\documentclass[letterpaper,twocolumn,10pt]{article}
\usepackage{usenix,epsfig,endnotes}

\usepackage[usenames,dvipsnames]{color}
\usepackage{hyperref}
\usepackage{subfigure}
\usepackage{amsmath}
\usepackage{pgfplots}
\usepackage{algpseudocode,algorithm,algorithmicx}
\usepackage{float}
\usepackage{multirow}

\usepackage{graphicx}
\usepackage{booktabs}
\usepackage{xparse}
\usepackage{tabularx}

\newfloat{algorithm}{t}{lop}

\usepgfplotslibrary{external} 
\NewDocumentCommand{\rot}{O{45} O{1em} m}{\makebox[#2][l]{\rotatebox{#1}{#3}}}%

\newenvironment{packed_itemize}{
\begin{itemize}
  \setlength{\itemsep}{1pt}
  \setlength{\parskip}{0pt}
  \setlength{\parsep}{0pt}
}{\end{itemize}}

\begin{document}
\date{}

\title{\Large \bf Hacking in the Blind:\\ 
(Almost) Invisible Runtime UI Attacks on Safety-Critical Terminals}

\author{
{\rm Luka\ Malisa}\\
Institute of Information Security\\
ETH Zurich
\and
{\rm Kari Kostiainen}\\
Institute of Information Security\\
ETH Zurich
\and
{\rm Thomas Knell}\\
ETH Zurich
\and
{\rm David Sommer}\\
Institute of Information Security\\
ETH Zurich
\and
{\rm Srdjan Capkun}\\
Institute of Information Security\\
ETH Zurich
} 


\maketitle


\subsection*{Abstract}

Many terminals are used in safety-critical operations in which humans, through terminal user interfaces, become a part of the system control loop (e.g., medical and industrial systems).  These terminals are typically embedded, single-purpose devices with restricted functionality, sometimes air-gapped and increasingly hardened.

We describe a new way of attacking such terminals in which an adversary has only temporary, non-invasive, physical access to the terminal. In this attack, the adversary attaches a small device to the interface that connects user input peripherals to the terminal. The device executes the attack when the authorized user is performing safety-critical operations, by modifying or blocking user input, or injecting new input events. 

Given that the attacker has access to user input, the execution of this attack
might seem trivial. However, to succeed, the attacker needs to overcome a
number of challenges including the inability to directly observe the user interface and avoid being detected by the users. We present techniques that allow user interface state and input tracking.  We evaluate these techniques and show that they can be implemented efficiently. We further evaluate the effectiveness of our attack through an online user study and find input modification attacks that are hard for the users to detect and would therefore lead to serious violations of the input integrity.  


\section{Introduction}
\label{sec:intro}

Many embedded terminals are used for safety-critical operations. For example, doctors program medical implants (e.g., Cardioverter Defibrillators (ICDs) and pacemakers) using dedicated terminals. At industrial facilities, the operation of automated devices, such as assembly robots and process control systems, are adjusted from control terminals. If the adversary manages to modify the operation of such terminals at a critical moment, the consequences can be severe.

\begin{figure}[t]
  \centering
  \includegraphics[width=\linewidth]{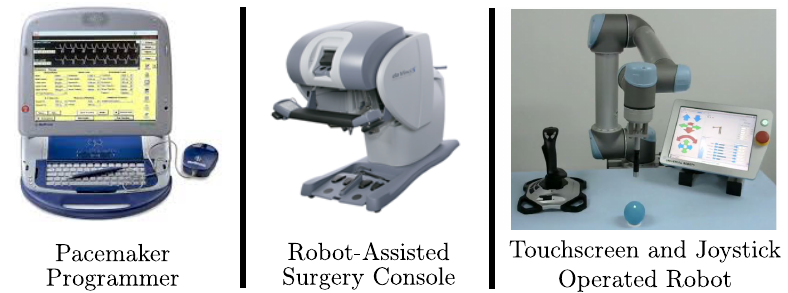}
  \vspace{-5mm}
  \caption{Examples of safety-critical terminals.}
  \label{fig:terminals}
\end{figure}

In contrast to general-purpose PC and smartphone platforms, these terminals are single-purpose devices with stripped-down functionality, and often hardened security. The terminals run operating system versions, such as Embedded Linux and Windows Embedded, that are smaller in size than general-purpose systems and can be configured to support relevant features only. The terminals are typically disconnected from the Internet and installation of third-party applications from external media is often not enabled. The Windows Embedded platform supports verification of the software configuration at boot and allows to be configured such that only signed software can be run (AppLocker feature) \cite{ms-lockdown}. The terminals typically run only a single UI application that is executed with least privileges. In such an environment, the user cannot modify the terminal settings besides what the application UI enables, in particular he cannot execute programs with administrative privileges. 

Attacking such hardened terminals can be more challenging compared to general-purpose systems. Remote attacks over the Internet might not be possible if the terminals are air-gapped. If the adversary obtains temporary physical access, he cannot easily install malware, as installation and execution of unsigned software may be prevented. A sophisticated attacker might be able to physically inject malicious code to the memory of the terminal at runtime, but such invasive attacks are elaborate to mount and may require expensive equipment. 

Another attack vector for such terminals is through their user interfaces. For example, many medical devices do not implement user authentication (password checks), as doctors must be able to operate them at all times. If the adversary can briefly access an unattended terminal, he can change its settings from the user interface. The attack is quick and non-invasive, but easier to notice and less severe compared to a runtime attack where the adversary modifies the operation of the terminal during its use (e.g., a doctor configuring an implant). Such attacks can also be prevented by mandating user authentication.

In this paper, we describe a novel way to attack safety-critical terminals through their user interface. Our approach violates the integrity of user input when the terminal is operated by the legitimate user, causing very damaging and stealthy runtime attacks without any malware running on the terminal.

In the attack, the adversary gains temporary physical access to a terminal, attaches a small attack device to it, and leaves the location. The attack device is attached to an interface that connects a user input device to the terminal. For example, the adversary can attach the attack device to a USB port that connects a mouse to the terminal or to an interface that connects an integrated touch screen to the terminal mainboard. The attack device observes the communication from the input device and when the legitimate user is performing a safety-critical operation, the device launches the attack by injecting new user input events. The attack can result in a serious safety violation, even loss of a human life, and the user is let to believe that he accidentally caused the damage himself.

The attack requires only brief and non-invasive access to the terminal. For example, attaching a small device to the USB port takes only a few seconds. A small attack device can be difficult to notice, and to a malware detection system the attack is invisible, as no malicious code is running on the terminal itself. The attack is also agnostic to any applied user authentication. The only chance to detect the attack is that the user notices the subtle visual changes on the user interface when the attack is active (e.g., medical device settings are modified). However, the attack can be made very fast and most people do not notice minor visual changes~\cite{simons2005}. 

Although the adversary has access to user input, realizing the attack involves technical challenges: 

\begin{packed_itemize}
\item Embedded terminals accept constrained user input and they do not allow admin access. 

\item The attack device operates ``blindly'' as it cannot observe the state of the UI or possible pointer location.

\item The users may notice input modifications or insertion, such as value change or pointer displacement.  
\end{packed_itemize}

To address them, we developed a novel state tracking algorithm that estimates the most likely system state and designed attack launch techniques that give little visual indication to the user. Our attack is invisible to traditional malware detection, it operates under a limited view of the target system without any feedback from the system, and it gives little visual indication to the user --- we call it \emph{hacking in the blind}.

We used the user interface of a medical implant programmer to evaluate our attack techniques, but we emphasize that our approach is applicable to a wide range of target systems from general-purpose PC platforms to hardened embedded terminals and different user input methods. Our evaluation shows that our algorithm can accurately determine the current state of the terminal. We tested UI manipulation techniques on 987 online study participants and noticed that our attacks can be very hard to detect: the attack success rate for the most stealthy variant was 93-96\%. We analyzed possible countermeasures and note that all of them have limitations. We conclude that our attack presents a serious threat to many embedded terminals.

To summarize, we make the following contributions: 

\begin{packed_itemize}
	\item \emph{New way to attack safety-critical terminals}. We propose an attack that is quick to deploy, hard for users to notice, and invisible to existing malware detection.

	\item \emph{New attack techniques.} We developed a novel algorithm that estimates the most likely state of the target system based on the observed user inputs and UI manipulation techniques that provide little visual indication to the user.

    \item \emph{Analysis of protective measures.} We analyze possible countermeasure and point out their limitations.
\end{packed_itemize}

The rest of the paper is organized as follows. In Section~\ref{sec:scenario} we explain the attack scenario. We describe our attack techniques in Section~\ref{sec:overview} and evaluate them in Section~\ref{sec:evaluation}. We analyze countermeasures in Section~\ref{sec:analysis}. Section~\ref{sec:discussion} provides discussion, Section~\ref{sec:related} reviews related work and Section~\ref{sec:conclusion} concludes the paper.

\section{Attack Scenario}
\label{sec:scenario}

We consider a scenario where the adversary has brief physical access to the target system. For example, in hospitals visitors can enter patient rooms where medical terminals are kept and in industrial facilities the cleaning personnel routinely has access to safety-critical control terminals.

\begin{figure}[t]
  \centering
  \includegraphics[width=\linewidth]{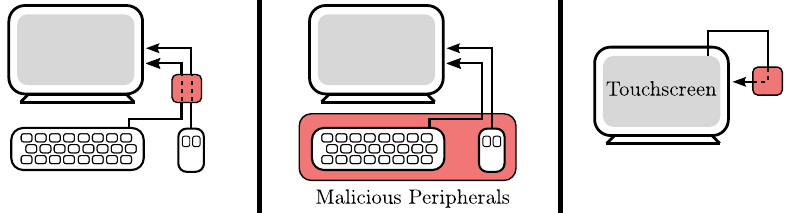}
  \vspace{-5mm}
  \caption{Attack scenario. The adversary attaches an attack device to an interface that connects a user input device to the terminal. The adversary can also replace a user input device with a malicious one.}
  \label{fig:attachment}
\end{figure}

In our attack, the adversary attaches a small attack device that sits in-between a user input device and the terminal (Figure~\ref{fig:attachment}). If the input device is integrated into the terminal device (e.g., touchscreen), the adversary attaches the attack device to an interface that connects the input device to the terminal mainboard. This may require opening the terminal enclosure. If the input device is an external device (e.g., USB mouse), the adversary can attach the attack device to the USB port that connects the device to the terminal. The adversary can also simply replace an external input device with one that contains the attack device. Such operations are quick to perform. 


The installed attack device observes user events from the connected input device and launches the attack by modifying user input or injecting new input events when the legitimate user is performing a safety-critical operation. Such attacks are the most severe in many scenarios. 

A likely outcome of the attack is that the user would believe that he caused the damage himself. Forensic analysis would also reveal a set of user inputs that lead to the accident, without any evidence of malware running on the device. Once the attack is done, the adversary can retrieve the attack device to remove evidence.





\textbf{Target system assumptions.} The target safety-critical system is an embedded terminal that runs a stripped-down operating system with hardened security. Here we describe protective measures (\emph{Lockdown features}) available on the Windows Embedded Industry platform \cite{ms-lockdown}, but similar security enhancements are commonly available on embedded Linux distributions as well. 

The terminal is a single-application system where the application user interface occupies the entire device screen. The terminal can be configured to run only one application (AppLocker feature). The user cannot escape the application UI with a specific key sequence (Keyboard Filter), and thus the user can only interact with the terminal through the application UI. The application is executed with least user privileges (User Account Control). We assume that the terminal is disconnected from the Internet. Installation of third-party software from external media is not allowed, and the terminal can verify its software configuration at boot and start only signed software at runtime (AppLocker). The terminal can be configured to only connect to USB devices with known class, device and product identifiers (USB Filter). Additionally, the entire terminal memory can be scanned for malware periodically. Side-channels, such as power consumption analysis, can be used to identify (malicious) processes running on the terminal \cite{clark-healthtech13}.

When the terminal is booted, its execution always begins from the same state. The application user interface is deterministic: similar interaction always causes the same result (e.g., state transition or remaining in the same state). The terminal is used via touchscreen, mouse or keyboard.

\textbf{Adversary capabilites.} We assume that the adversary can physically install the attack device unnoticed. Besides installing the attack device, the adversary does not interact with the terminal in any other way. In particular, we assume that the adversary does not reboot the terminal and that the adversary cannot observe its current state (the user interface of some terminals might be locked and password protected). The adversary can make the attack device so small that legitimate users do not notice its presence. If the device is used via two input devices (e.g., mouse and keyboard), the adversary can connect both of them to the same attack device. The attack device can observe, delay, and block all events from the connected user input devices as well as inject new events. After the installation, the attack device does not communicate with the remote adversary. After the attack, the adversary may collect the attack device. The adversary knows the user interface of the terminal, including its states and state transitions.


\section{Hacking in the Blind}
\label{sec:overview}

In our attack scenario, the adversary gains temporary access to the target terminal, attaches the attack device to it and leaves the location. The next time when the legitimate user operates the terminal, the attack device starts observing user input events. The attack device can intercept all events, but their interpretation may have two forms of uncertainty. 

First, the adversary may not know the state of the terminal user interface, e.g., because the user interface of the terminal was locked when the attack device was installed. We call this \emph{state uncertainty}. Second, the adversary may not be able to interpret all received user input events without ambiguity. In particular, mouse events are relative to the mouse cursor location that may be unknown to the adversary. We call this \emph{location uncertainty}. In contrast to mouse input, touchscreen events do not have location uncertainty.

The goal of the adversary is to attack the terminal through its user interface when the legitimate user is performing a safety-critical operation. The primary challenge is to launch the attack accurately under such uncertainty, without any feedback from the target terminal (hacking in the blind). The best attack strategy depends on the usage of the terminal, the type of the input device, the user interface configuration, and the level of stealthiness the adversary wants to achieve. We explore attack strategies from simple techniques and to more challenging scenarios that require more sophisticated solutions.

\subsection{Simple Techniques}

If the adversary manages to reduce (or remove) both location and state uncertainty, attacking the terminal user interface becomes easier. Assuming that the adversary knows both the current user interface state and the mouse cursor location, each received event can be interpreted unambiguously. An adversary that knows the user interface can easily track both mouse movement and state transitions in the user interface. Below we list methods that can help the adversary to reduce uncertainty.

\textbf{Reducing state uncertainty.} A simple technique to learn the state of the system is to wait for a reboot. If the attack device can determine when the terminal is booted, it knows that, shortly after, the terminal user interface is in a known state. This technique works only if the target terminal is rebooted before the attack. While some safety-critical terminals may be shut down after each use, others may run long periods of time without reboots. 

The technique has also another possible limitation. The user might create input events, such as mouse clicks, during the terminal boot process. Depending on the implementation of the terminal, the attack device may not know when exactly the user input device driver has been fully initialized, the application is started, and the terminal OS starts passing the incoming events to the application user interface. If the user clicks the mouse during the boot, the adversary might not be able to determine if the clicks reached the application and caused a state transition in its user interface. The adversary could address this by blocking events during boot.

Another simple technique is to wait for specific input event sequences. For example, if the terminal user interface has an editable text field in one state only, text input received from the keyboard is an indication that the terminal user interface is currently in that state. 

This technique has two limitations. First, in many user interfaces there are no user input event sequences that identify the state precisely (e.g., editable text fields in several states). Second, even if the user interface configuration has an event sequences that identifies the state, the sequence may not manifest in every usage of the terminal (e.g., no text input every time).

\textbf{Reducing location uncertainty.} A simple technique to determine the mouse cursor location is to actively move the mouse (i.e., inject movement events) towards a corner of the screen. For example, if the mouse is moved up and left sufficiently, the adversary knows with certainty that the mouse cursor is located at the top-left corner of the screen. Moving the mouse while the system is idle may not be possible, if the terminal user interface is locked. To make the above process appear less suspicious during terminal use, the adversary can create an appearance that the user moved the mouse herself. For example, when the attack device observes mouse movement events left and up, it can inject additional movement events to the same direction. The process mimics a situation where the mouse movement was accelerated and the user unintentionally moved the mouse cursor to corner of the screen herself. The limitation of this technique is that if such mouse movement is performed repeatedly (e.g., at every boot), it can appear suspicious to an anomaly detection system or post-attack forensics. 

Waiting for a reboot is another possible way to learn the mouse location. Typically, the mouse cursor is placed at the same location on the screen after boot. The above discussed limitations apply also to this technique. The user might move the mouse during the boot process which makes it harder to tell where the mouse cursor is after the boot.

\begin{figure}[t]
  \centering
  \includegraphics[width=\linewidth]{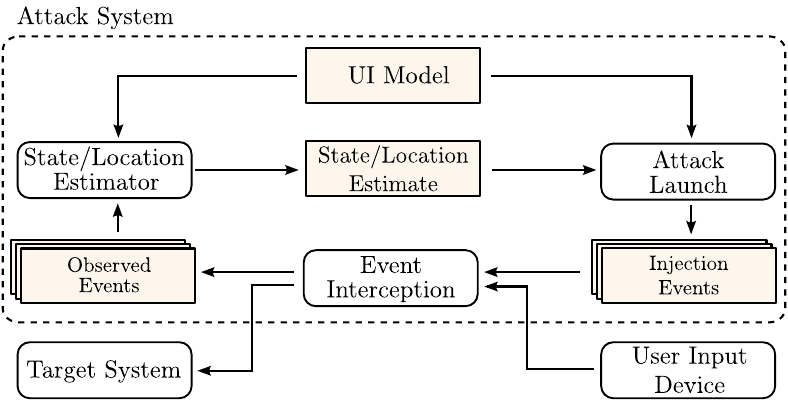}
  \vspace{-5mm}
  \caption{Attack system overview. On the attack device, a state tracking component processes observed user input events with respect to a UI model and produces a state estimate. The attack is launched based on the state estimate and the UI model by blocking and injecting new events.}
  \label{fig:overview}
\end{figure}


We take a practical stand and assume that in many scenarios the adversary has to perform the attack under location uncertainty, state uncertainty or both. The case with both location and state uncertainty is the most challenging to attack and, to us, the most interesting.

\subsection{Attacks Under Uncertainty}

Starting from this section, we describe a novel attack system that enables the adversary to launch accurate attacks despite of uncertainty. A noteworthy property of the system is that it estimates user interface state and mouse location fully passively, and thus enables implementation of stealthy and precise attacks. We proceed by giving a high-level overview of the attack system (Figure~\ref{sec:overview}). 

The attack device contains a static model of the target system user interface that the adversary constructed before the device deployment and it runs two main software components. The first component is a \emph{State and Location Estimator} that determines the most likely user interface state (and mouse cursor location) based on the observed user events and the UI model. The estimation process can begin from a known, or an unknown user interface state, at an arbitrary moment and it tracks mouse, keyboard and touchscreen events. The second component is an \emph{Attack Launcher} that performs active UI manipulation when the legitimate user is performing a safety-critical operation. We describe several attack variants and evaluate their detection through a user study (Section~\ref{sec:evaluation}).

\subsection{User Interface Model} 

\begin{figure}[t]
  \centering
  \includegraphics[width=\linewidth]{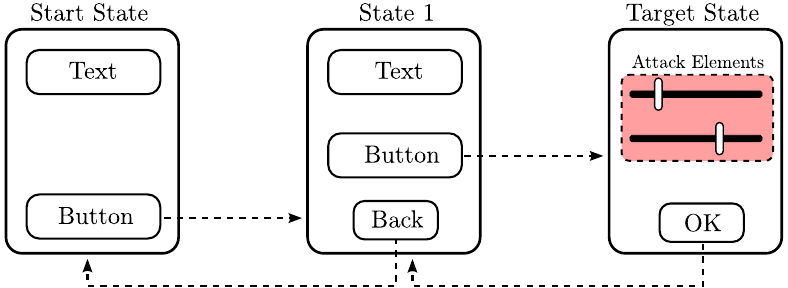}
  \vspace{-5mm}
  \caption{Example user interface model. The model includes user interface states, user input elements and state transitions. The attack state contains attack elements and a confirmation element.}
  \label{fig:model}
\end{figure}

The user interface model (see Figure~\ref{fig:model}) contains user interface states, their user input elements and state transitions. User input elements are buttons, editable text fields, multiple choice elements, movable sliders etc. All input elements can be interacted with mouse and touchscreen devices. Some user input elements can be interacted with a keyboard device. 

For each state, the model includes the locations and the types of the user input elements and the possible state transition that the element triggers. The transitions are deterministic. One of the states is defined as the start state and one or more states are defined as the target states. The goal of the attack is to modify safety-critical input elements (\emph{target elements}) on the target states. Typically the target state includes also a confirmation element that the user clicks to confirm the safety-critical operation. 

\subsection{State and Location Estimation} 

\begin{figure}[t]
  \centering
  \includegraphics[width=\linewidth]{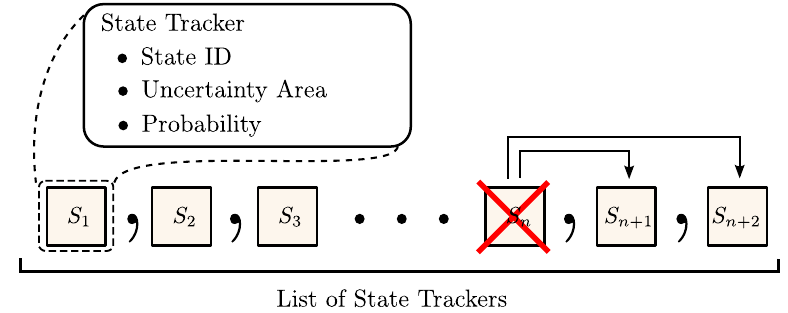}
  \vspace{-5mm}
  \caption{Our algorithm maintains a list of state trackers. On each click it creates child trackers for every tracker and removes the parent from the list.}
  \label{fig:trackers}
\end{figure}

Here we describe our state and location estimation algorithm for mouse and keyboard (both location and state uncertainty). Later we explain how the same algorithm can be used to estimate state for touchscreen input (only state uncertainty).

The algorithm operates by keeping track of all possible user interface state and mouse location combinations. For each possible state and location the algorithm maintains a \emph{state tracker} object. The state trackers contain an identifier of the state and an \emph{uncertainty area} that determines the possible location of the mouse in that state instance. Additionally, the algorithm assigns a probability for each tracker object that represent the likelihood that the terminal user interface and the mouse cursor are in this state and location.

The estimation algorithm maintains the tracker objects in a list (Figure~\ref{fig:trackers}). If the estimation begins from a known state, we have initially only one tracker in the list to which we assign 100\% probability. If the estimation begins from an unknown state, we create one tracker per possible system state and assign them equal probabilities. Assuming no prior knowledge on the mouse location, we set the mouse uncertainty area to cover the entire screen in each tracker during initialization.

The state and location estimation is an event-driven process. Based on the received user input events, we update the trackers on the list, create new trackers and delete trackers from the list. For each mouse movement event, we update the mouse uncertainty area in each tracker. For every mouse click, we consider all possible outcomes of the click, including transitions to new states, as well as remaining in the same state. We create new \emph{child trackers} with updated uncertainty areas, add the children to the list, and remove the \emph{parent tracker} from the list (see Figure~\ref{fig:trackers}). When we observe a user event sequence that indicates interaction with a specific UI element, we update the probabilities of each tracker accordingly. We explain these steps in detail below. 

\begin{figure}[t]
  \centering
  \includegraphics[width=\linewidth]{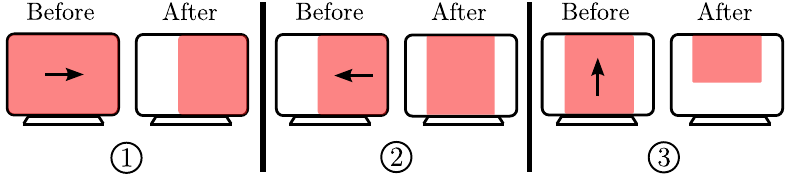}
  \vspace{-5mm}
  \caption{Movement event handling. Movement can reduce the uncertainty area size (1) and (3) or change its location (2).}
  \label{fig:area}
\end{figure}

\textbf{Movement event handling.} When the mouse uncertainty area is the entire device screen, any mouse movement reduces the size of the uncertainty area. For example, if the user moves the mouse to the right, the area becomes smaller, as the mouse cursor can no longer reside in the leftmost part of the screen (Figure~\ref{fig:area}). If the mouse is moved to a direction where the uncertainty area border is not on the edge of the screen, the mouse movement does not reduce the size of the uncertainty area, but only causes its location to be updated. Any mouse movement towards a direction where the uncertainty area is on the border of the screen, reduces the size of the uncertainty area further. For each received mouse movement event, we update the uncertainty areas in all trackers.

\begin{figure}[t]
  \centering
  \includegraphics[width=\linewidth]{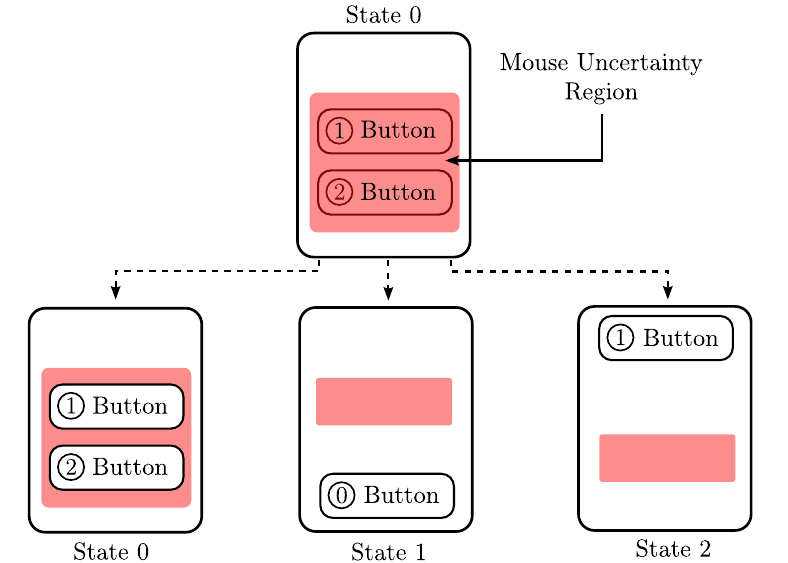}
  \vspace{-5mm}
  \caption{Click event handling. We create state trackers for all possible click outcomes, including remaining in the same state and transitions to new states. A new uncertainty area is calculated for each tracker.}
  \label{fig:tree}
\end{figure}

\textbf{Click event handling.} When we observe a mouse click event, the estimation algorithm considers all possible outcomes for each tracker. The possible outcomes are determined by the current mouse uncertainty area (Figure~\ref{fig:tree}). For each possible outcome we create new child trackers and update their mouse uncertainty areas as follows. 

If the user interface remains in the same state, the updated mouse area for the child is the original area of the parent from which we remove the areas of the user input elements that cause transitions to other states. For each state transition, the mouse area is calculated as the intersection of the parent area and the area of the user input element that caused the transition. Once the updated mouse uncertainty areas are calculated for each child tracker, we remove the parent tracker from the list, and add the children to it. We repeat the same process for each state tracker on the list. We note that as a result of this process, the list may contain multiple trackers for the same state with different mouse uncertainty areas.

\begin{figure}[t]
  \centering
  \includegraphics[width=\linewidth]{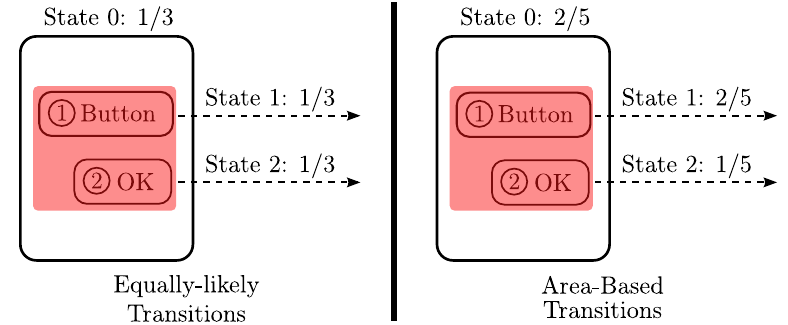}
  \vspace{-5mm}
  \caption{Transition probabilities. On the left, all possible outcomes are considered equally like. On the right, we illustrate area-based transition probabilities.}
  \label{fig:prob}
\end{figure}

The probability of a child tracker is calculated by multiplying the probability of its parent with a \emph{transition probability}. We consider two options for assigning transition probabilities, as shown in Figure~\ref{fig:prob}.

\begin{packed_itemize}
	\item \emph{Equal transitions.} Our first option is to consider all possible state transitions equally likely. E.g., if the mouse uncertainty area contains two buttons, each of them causing a separate state transition, and parts of the screen where a click does not cause a state transition, we assign each of them 1/3 probability.

	\item \emph{Element area.} Our second option is to calculate the transition probabilities based on the surface of the user interface element covered by the mouse uncertainty area. For example, if the uncertainty area covers a larger area over one button than another, we assign it bigger transition probability.
\end{packed_itemize}

The transition probabilities can be enhanced with \emph{a priori probabilities} of UI element interactions. For example, based on prior experience on comparable user interfaces, the adversary can estimate that an OK button is pressed twice as likely as a cancel button in a given state.  

\textbf{Element detection.} Finally, we identify user interaction with certain UI elements based on sequences of observed user input events. For example, a mouse event sequence that begins with a button down event, followed by movement left or right that exceeds a given threshold, followed by a button up event is an indication of slider usage. Similarly, text input from the keyboard indicates likely interaction with an editable text field and a click indicates a likely interaction with a button.

When we observe such event sequences (slider movement, text input, button click), we update the probabilities of the possible trackers on the list. One possible approach would be to remove all trackers from the list where interaction with the identified element is not possible (e.g., a button click is not possible under the mouse uncertainty area). After such trackers would be removed from the list, we could increase the probabilities of the remaining ones equally. Such an approach could yield fast results, but also provide erroneous state estimations. If the user provides text input on a user interface state that does not contain editable text fields or if text highlighting is mistaken for slider movement, the algorithm would remove the correct state from the list. 

We adopt a safer but slower approach where we consider trackers with the identified elements more likely and scale up their probabilities, and keep the remaining trackers and scale down their probabilities. The scaling factor is an adjustable parameter of this approach. 

\textbf{Target state detection.} Our algorithm continues the state tracking process until two criteria are met. First, we have identified the target state with a probability that exceeds a given threshold. After each click event and detected element we sum the probabilities for all trackers that represent the same state to check if any of them exceeds the threshold. Second, the mouse uncertainty area must be small enough to launch the attack. We combine the mouse uncertainty areas from all matching trackers and consider the uncertainty area sufficiently small when its size is smaller than the size of the target elements or the confirmation element.

\textbf{State estimation for touchscreen.} Using a touchscreen instead of a mouse does not affect our algorithm. Typically touchscreens report click events in absolute coordinates, hence using a touchscreen corresponds to the case where the mouse location is known, but the starting state is not. Determining the possible transitions after a click is trivial, since there can be at most one intersection of a clicking point with the area of an element in a specific state. Furthermore text input can be observed from the virtual keyboard therefore the element detection works the same way as described previously.

\subsection{Attack Launch Techniques}
\label{sec:launch}

Once the attack device has identified the attack state with sufficiently small uncertainty area, it is ready to launch the attack. In a simple approach, the adversary moves the mouse cursor over one of the attack elements, modifies its value, moves the mouse cursor over the confirmation button, and clicks it. The process is fast and the user has little chances of preventing the attack. However, the user is likely to notice such an attack. For example, if a doctor never clicked the confirm button herself, she is unlikely to implant the programmed pacemaker into a patient. For this reason, we focus on more subtle attack launch techniques. Below we describe two such techniques and in Section~\ref{sec:evaluation} we evaluate their user detection.

\textbf{Element-driven attack.} The adversary first identifies that the user interacts with one of the target elements. This can be easily done when the mouse uncertainty area is smaller than the target element. Once the user has modified the value of the target element, the adversary waits a small period of time and during it tracks the mouse movement, then quickly moves the mouse cursor back to the target element, modifies its value, and returns the mouse cursor to its location. After that, the adversary lets the legitimate user confirm the safety-critical operation. The technique only requires little mouse movement, but the modified value remains visible to the user for a potentially long time, as the adversary does not know when the user will confirm the safety-critical operation. 

\textbf{Confirmation-driven attack.} The adversary identifies that the system is on the attack state and lets the user to set the attack element values uninterrupted. When the user clicks the confirmation button, the attack activates. The adversary blocks the incoming click event, moves the mouse cursor over one of the attack elements, modifies its value, moves the mouse cursor back over the confirmation button, and then passes the click event to the target system. After that, the adversary changes the modified attack element back to its original value. In this technique, the mouse cursor may have to be moved more, but the modified attack element settings remain visible to the user only a very short period of time.

\section{Evaluation}
\label{sec:evaluation}

In this section we evaluate state and location estimation performance and report results from an online user study where we tested how many users would detect our attacks. 

\subsection{Evaluation Setup}
\label{sec:case}

To evaluate the estimator performance and attack detection we built an application that simulates the user interface of a pacemaker programmer based on publicly available documentation of an existing cardiac implant programmer~\cite{pacemaker_programmer}. Such a programmer terminal is used by doctors to configure medical implant settings. For example, when a doctor prepares a pacemaker for implantation, she configures its settings based on the heart condition of the receiving patient. The terminal can also be used to monitor the operation of the implant and potentially update its settings. The user interface was designed for mouse and keyboard use. 

\begin{figure}[t]
  \centering
  \includegraphics[width=\linewidth]{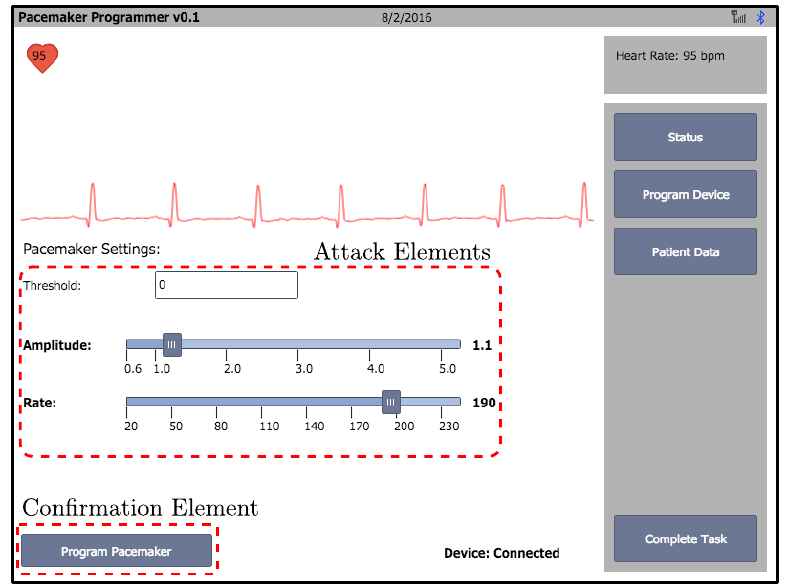}
  \vspace{-5mm}
  \caption{Our custom cardiac implant programmer screenshot. The attack and the confirmation elements are highlighted.}
  \label{fig:screenshot}
\end{figure}

The model of this user interface consists of approximately ten states and contain three types of user input elements: buttons, text fields and sliders. All state transitions are triggered by button clicks. The attack elements are the user input elements that are used to configure the pacemaker settings. Threshold is set using a text field (keyboard), while amplitude and rate are set using slider elements (mouse), see Figure~\ref{fig:screenshot}. All attack elements are on the same state. The model creation was a manual process that took us a few hours. 

\begin{figure}[t]
  \centering
  \includegraphics[width=\linewidth]{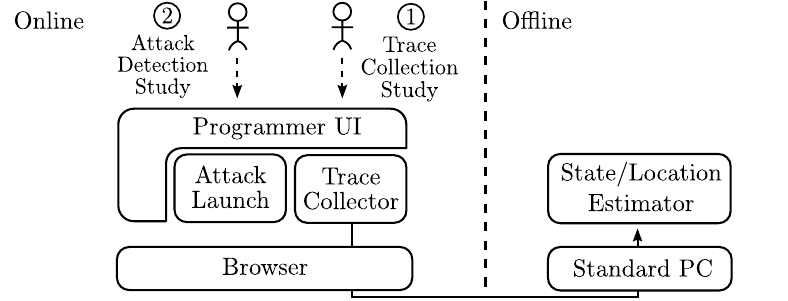}
  \vspace{-5mm}
  \caption{Our evaluation setup. We collected user traces and tested attack detection online. We evaluated estimator performance on separate implementation offline.}
  \label{fig:eval-setup}
\end{figure}

\textbf{Evaluation setup.} On a real attack scenario, both the state estimator and the attack launcher would be implemented inside the same attack device (e.g., small dongle). To demonstrate the feasibility of hacking in the blind, we implemented the state estimator and the attack launching as two separate components(Figure~\ref{fig:eval-setup}). Our programmer user interface was implemented as a web application. We used it to collect user traces online. We evaluated the collected traces on the state estimator offline. Attacks were tested on the browser platform. 

We implemented the state and location estimator using Python and the pacemaker programmer UI using Haxe language that was compiled to the HTML5 (JavaScript) backend.

\subsection{Trace Collection}

To evaluate the tracking algorithm we collected user traces for the programmer user interface online.

\textbf{Participant recruitment.} We recruited 400 participants for trace collection using a crowd sourcing platform called CrowdFlower. The platform enables the definition of typically small online jobs that human contributors complete in return of a small payment. We recruited participants globally and required them to be at least 18 years old. Each contributor was allowed to complete only one job for trace collection. On CrowdFlower platform our job had a title \emph{``Program an implanted pacemaker''} and its description stated: 

\begin{quote}
We are evaluating the user interface of an experimental medical device. Your task is to configure a pacemaker device by interacting with the pacemaker programming software. Note that this is a test! The shown user interface is not connected to a real patient.
\end{quote}

\textbf{Task details.} In each job, we asked the participants to fill in a short questionnaire that we used to collect demographic information. The questionnaire included also a test question, with a known answer, that we used to filter out participants that were clearly not attentive. After the questionnaire, the participants were shown more detailed task instructions and the pacemaker programmer user interface. The participants interacted with the user interface using a mouse and a keyboard on their browsers. 

In the instructions we asked the participant to find saved patient data that matches a given medical condition, copy that patient's pacemaker settings to the programming screen, and finally, to program the device by pressing the confirmation element. The full instructions are listed in Appendix~\ref{sec:supplementary}. They remained visible to the participant while she interacted with the programmer UI. We recorded all user input during the task, but no private information on study participants was collected.


\begin{figure*}[ht]
 	\centering
	\includegraphics[width=0.9\textwidth]{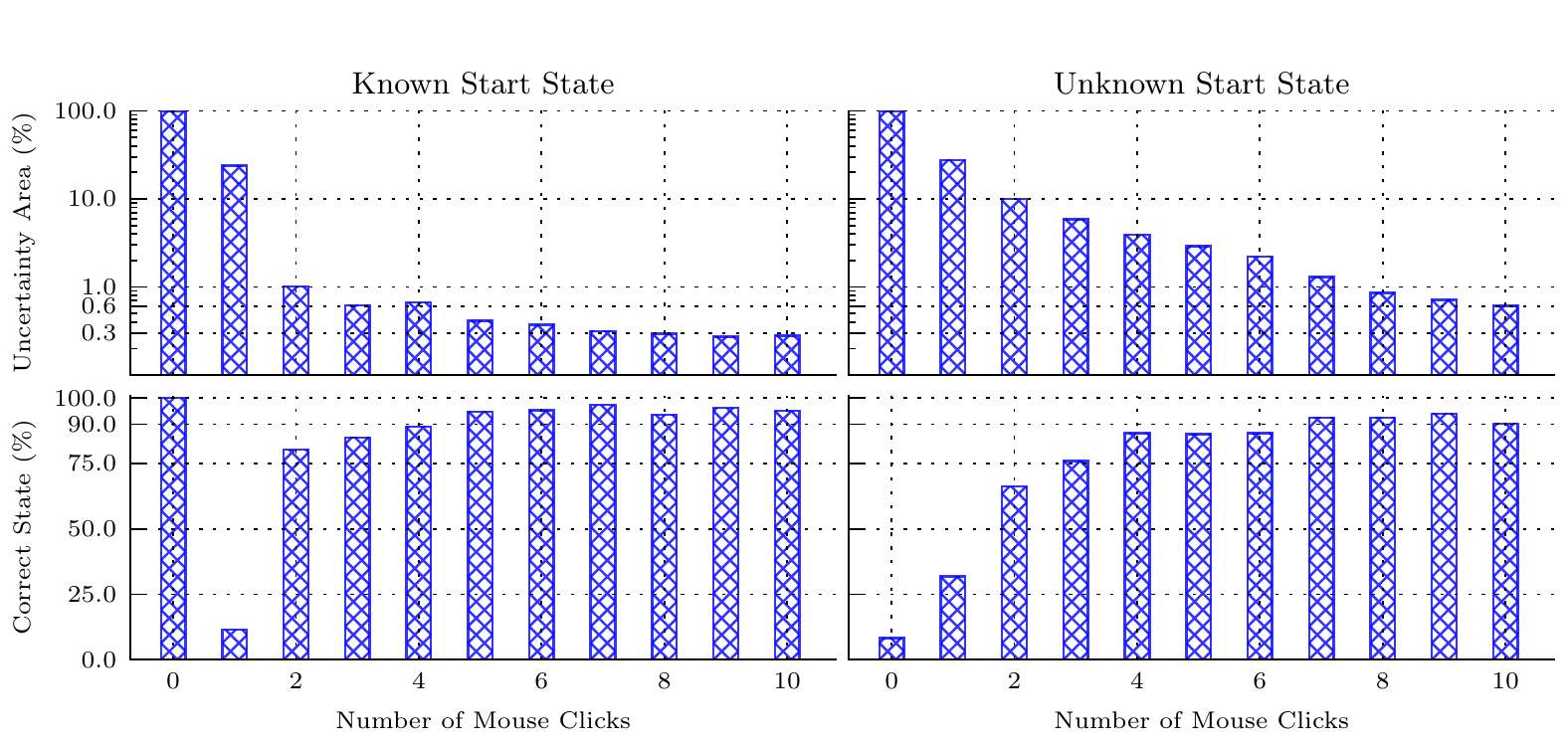}
	\vspace{-5mm}
	\caption{State tracking accuracy. On the left, tracking from a known state. The uncertainty area reduces fast. After ten clicks, we estimate the correct state with high probability. On the right, tracking from an unknown start state. The uncertainty area reduces slower and we estimate the correct state with slightly lower probability. }
	\label{fig:uncert_state}
\end{figure*}

\textbf{Trace analysis.} In total 400 contributors completed the task, Table~\ref{tab:trace-demo} in Appendix~\ref{sec:supplementary} shows their demographics. We divided the collected traces randomly into 200 training traces and 200 evaluation traces. We analyzed all the training traces. The time required to complete the task varied greatly. On the average the traces had 29 ($\pm 22$) clicks and 98\% of the traces had at least ten clicks. We profiled each user interface state and calculated how often each button was pressed. We analyzed also conditional button press frequencies, i.e., how often a button was pressed given that the user transitioned to the current state from a given previous state. By analyzing the traces, we observed that approximately 7\% of user input gestures were over wrong or non-existent elements. For example, users clicked when the mouse cursor was not over a button element. 



\subsection{Estimation Accuracy}

We ran our estimation algorithm (implemented in Python) on all our evaluation traces. As our algorithm is event-based, after each click we measured (a) the size of the mouse uncertainty area, expressed as the percentage of the overall screen size, and (b) the probability that we correctly estimate the real state the user was currently in. Figure~\ref{fig:uncert_state} shows our results.

We say that our algorithm correctly estimates the current state, when it assigns the highest probability for the correct state among all states. As all our traces start from the same state, to evaluate the situation where the tracking begins from an unknown state, we cut the first 10\% from all our evaluation traces. As tracking options we used the element-area transition probabilities together with element detection (scaling parameter 0.95) and a priori probabilities that we obtained by profiling the training traces.  

First, we discuss the case where the state tracking begins from a known start state (shown left in Figure~\ref{fig:uncert_state}). The uncertainty area is the full screen at first and the probability for estimating the correct state is 100\% (known start state). As the estimation algorithm gathers more user input events, the uncertainty area size reduces quickly and already after three clicks the area is less than 1\% of the screen size. The estimation probability decreases first, as the first click adds uncertainty to the tracking process, but after additional click events, the estimation probability increases steadily, and after ten clicks the algorithm can estimate the correct state with above 90\% probability.

Next, we consider the scenario where the state tracking begins from an unknown target system state (shown right in Figure~\ref{fig:uncert_state}). In the beginning, the uncertainty area is the entire screen and the probability for the state estimate is low, as all states are equally likely. As the tracking algorithm gathers more user events, the uncertainty area reduces, but not as fast as in the case of known start state. The uncertainty area becomes less than 1\% of the screen size after eight clicks. The probability for the correct state estimate increases and after ten clicks we can estimate the correct state with approximately 90\% probability. We do not report uncertainty area and state probability past ten clicks, as many of our traces were not longer than that.

We conclude that in both cases we can identify the correct system state with high probability after observing only ten clicks and the uncertainty area becomes very small (below 1\%, equal to a small, $50\times50$ pixel rectangle). In our case study user interface the target and the confirmation elements are significantly larger than 1\% of the screen size. If the user enters the attack state after ten clicks, we can launch the attack accurately. This was true in 96\% of our traces.

\textbf{Tracking option comparison.} We compared the performance of our different tracking options. We evaluated two transaction probability assignment schemes (equal transactions and element area) and tested both schemes with and without element detection and a priori probabilities. We measured the probability that we estimate the correct state after ten clicks, Table~\ref{tab:evaluation} shows the results. We notice that the equal transitions option performed slightly better than the area based probability assignment. Element detection gives a major detection accuracy improvement. A priori probabilities do not improve accuracy significantly. The correct state probability is higher when we analyze traces with unknown start state, but the mouse uncertainty is significantly larger than in the known starting state samples.

\begin{table}[t]
	\centering
	\scriptsize
	\begin{tabular}{llrrr}
	\hline
	                                     &              & \textbf{\scriptsize\shortstack{Transition\\probability}} & \textbf{\scriptsize\shortstack{+element\\detection}} & \textbf{\scriptsize\shortstack{+a priori\\probability}} \\
	\hline
	\multirow{3}{*}{\textbf{Known} \textbf{state}}   & equal trans. &34\% & 96\% & 94\% \\
                                                                 & element area &27\% & 96\% & 95\% \\
	\hline
	\multirow{3}{*}{\textbf{Unknown} \textbf{state}} & equal trans. &52\% & 90\% & 91\% \\
	                                                             & element area &44\% & 90\% & 90\% \\
	\hline
	\end{tabular}
	\caption{Tracking option comparison. We report the correct state probability after ten clicks in each option.}
	\label{tab:evaluation}
\end{table}


\textbf{Touchscreen tracking.} We also tested how accurately our algorithm could detect the correct state on a touchscreen scenario that has state uncertainty, but no location uncertainty. We used our evaluation traces that were collected from mouse and keyboard input, neglected the movement information and enhanced the click and drag events with absolute coordinates. We kept the text input, assuming that the adversary can collect text input from touchscreen virtual keyboard. We processed the modified traces using the equal transitions option, a priori probabilities and element detection starting the tracking from an unknown state. After five clicks, we were able to detect the correct state with 99\% probability.

\subsection{Estimation Overhead} 

\begin{figure}[t]
  \centering
  \includegraphics[width=\linewidth]{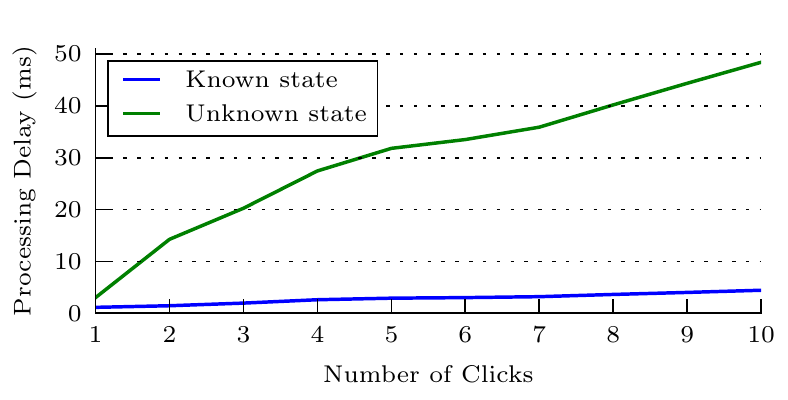}
  \vspace{-5mm}
  \caption{State tracking overhead. The tracking overhead per user input event increases over time as the state tracking algorithm accumulates more trackers.}
  \label{fig:performance}
\end{figure}

To analyze how fast our state and location estimation algorithm runs, we measured the runtime overhead of processing each user input event. Using the acquired traces, our algorithm (implemented in Python) was run on a standard desktop (Intel iCore 5, 3.4 GHz, 16 GB RAM) with all the tracking options enabled using element area transition probabilities. The \emph{processing delay} is the time our algorithm takes to process a single mouse click event, and this processing delay is related to the number of state trackers that our algorithm maintains. 

Figure~\ref{fig:performance} shows the results. When we start tracking from a known state, the overhead increases slightly over time, but remains under 10 milliseconds per event. When we start tracking from an unknown state, the algorithm accumulates significantly more trackers, and thus the processing overhead increases faster. After ten clicks, processing a single input event takes approximately 50 milliseconds on our test platform.

We conclude that the performance requirements of the tracking algorithm are moderate for simple user interfaces where the operations are typically completed with small number of clicks. In Section~\ref{sec:discussion} we discuss the performance requirements and possible optimization techniques for more complex UIs and longer tracking sessions.

\subsection{Attack Detection User Study}

To evaluate how many users would detect ours attacks, we conducted a second user study online. 

\textbf{Recruitment and procedure.} We created a new job on the same crowd sourcing platform with similar description and recruited 1200 new study participants. We divided the participants into 12 equally large attack groups of 100 participants each. We tested two element-driven attack variants: one where we modify a text input element and another we we modify a slider input element. We also tested two confirmation-driven attack variants: one with text and another with slider input. For each four attack variant we tested three separate speeds of the attack. We demonstrate the confirmation-driven attack variant that modifies a slider at 10 ms speed in a video that is available online.\footnote{{\small\url{https://goo.gl/rzCX3o}}}
 
For all participants, we provided the same task description as before, but depending on the group, we launched an attack during the task. Once the task was over, we asked the participants: ``Do you think you programmed the pacemaker correctly?'' with yes/no answer options. We also asked the participants to give freeform feedback on the task. 

If a participant noticed the UI manipulation, she had three possible ways to act on it. First, the participant was able to program the pacemaker again with the correct values. Second, the participant could report that the device was not programmed correctly in the post-test question. Third, the participant could write to the freeform feedback that she noticed something suspicious in the application user interface.

\textbf{Study results.} In total 987 participants completed the task and we report their demographics in Table~\ref{tab:attack-demo} (Appendix~\ref{sec:supplementary}). We consider that the attack succeeded when the participant did none of the above mentioned three actions. The results are shown in Table~\ref{tab:attack-results}. The success rate for the element-driven text attacks was 37-50\% and for the element-driven slider attacks 6-12\%, depending on the speed of the attack. The success rate for the confirmation-driven text attacks was 93-96\% and for the confirmation-driven slider attacks 90-95\%. 

\begin{table}[t]
  \centering
  \footnotesize
  \begin{tabular}{lcc}
    \hline
    \textbf{\shortstack{Attack\\group}} & \textbf{\shortstack{Attack\\succeeded}} & \textbf{\shortstack{Task\\completed}} \\
	\hline
	1. Element, text, 5 ms & 50\% & 84 \\
	2. Element, text, 62 ms & 37\% & 84 \\
	3. Element, text, 125 ms & 48\% & 86 \\
	\hline
	4. Element, slider, 5 ms & 12\% & 80 \\
	5. Element, slider, 62 ms & 9\% & 83 \\
	6. Element, slider, 125 ms & 6\% & 86 \\
	\hline
	7. Confirmation, text, 10 ms & 93\% & 81 \\
	8. Confirmation, text, 125 ms & 96\% & 79 \\
	9. Confirmation, text, 250 ms & 93\% & 78 \\
	\hline
	10. Confirmation, slider, 10 ms & 95\% & 85 \\	
	11. Confirmation, slider, 125 ms & 90\% & 82 \\	
	12. Confirmation, slider, 250 ms & 95\% & 79 \\    
	\hline
	\textbf{Total} & & \textbf{987} \\
	\hline 
  \end{tabular}
  \caption{Attack detection study results. For each attack group we report the the percentage of users against which the attack succeeded, and the number of participants that completed the task.}
  \label{tab:attack-results}
\end{table}

To compare the different types of attacks, we performed chi-squared tests of independence. First, we compared all the element-driven attacks to confirmation-driven attacks. As can be clearly seen from the results, the success rate in the confirmation-driven attacks was significantly higher ($\chi^2(1,N=987)=450.7), p<0.001$). Then we compared the attacks that modify text value to attacks that modify a slider value. In the element-driven attacks, text field manipulation had a higher success rate than slider movement ($\chi^2(1,N=503)=78.1), p<0.001$), but in the confirmation-driven attacks the type of the manipulated user interface element had no effect on the attack success rate ($\chi^2(1,N=484)=0.12), p=0.73$). 

Finally, we tested the effect of the attack speed and found that in none of the tested four attack types, the speed of the attack had a significant effect on the success rate. We report the test values for the element-driven text attacks ($\chi^2(2,N=254)=3.56), p=0.17$), element-driven slider attacks ($\chi^2(2,N=249)=1.21), p=0.54$), confirmation-driven text attacks ($\chi^2(1,N=238)=0.42), p=0.81$), and confirmation-driven slider attacks ($\chi^2(2,N=246)=2.20), p=0.33$).

We conclude that all the tested confirmation-driven attacks are very stealthy, in each variant the success rate was at least 90\%. In the element-driven attacks that user interface manipulation remains visible longer for the user, and this is a possible explanation why the attacks do not succeed equally well. However, our user study was not designed to prove or reject such hypothesis. Interestingly, we notice that in the confirmation-driven attacks, the attack variants of speed 250 milliseconds did not have significantly lower success rates than the faster attacks. This implies that the adversary has at least a few hundred milliseconds time to perform the user interface manipulation without sacrificing its success rate.

\textbf{Discussion.} We analyzed the freeform text responses from all users, and none of the users associated the observed UI changes to a malicious attack. Only two users commented on the changing values of UI elements and both attributed the changes to a software glitch. One user noted \emph{``Possible bug when working with sliders. Threshold value changed from 88 back to 80, had to correct''}. This result shows that users are habituated to software errors in application user interfaces.

Out of the 987 study participants only 21 answered negatively to the question \emph{``Do you think you programmed the pacemaker correctly?''}. A possible explanation is that users misinterpreted the results of positive UI feedback. To reduce chances of errors, it is common for safety-critical systems to have strong positive feedback mechanisms (e.g., clearly visible user action notifications). Even though an attack was performed, the users could have been fooled into believing that nothing out of the ordinary happened by the benign and reassuring nature of our ``Device Programmed'' notification. A user remarked: \emph{``I was told at the end that the pacemaker was programmed, so I assume I did it correctly''}.

\section{Countermeasures}
\label{sec:analysis}


\textbf{Trusted input devices.} One way to address our attacks is to mandate usage of trusted input devices. We call a user input device \emph{trusted}, when it securely shares a key with the target system. For example, USB input devices communicate using polling. The host sends periodic requests and the input device sends responses that report a possibly occurred user event. With a shared key all request and responses can be encrypted and authenticated which prevents the adversary from observing and injecting events. If the responses also include a freshness guarantee, such as a nonce, the adversary cannot replay events either. 

However, secure deployment of trusted input devices is challenging. Assuming that the target system and the input device have a certified key, and the two devices run a mutually authenticated key agreement protocol at connection establishment. If the certified input device is temporarily unavailable (e.g., lost or broken), the safety-critical terminal cannot be operated with another, non-certified device. For example, doctors need  be able to operate medical terminals at all times. Additionally, the adversary can purchase a certified input device, extract a key from it, and install it to the attack device. Standard approaches to address compromised keys, such as online revocation checks, are ill-suited to this setting, as many safety-critical embedded terminals are disconnected from the Internet to limit their attack surface.

\textbf{Increased user feedback.} The user interface can provide visual feedback on each change on attack elements \cite{huang-usenix12}. For example, the user interface can draw a thick border around a recently edited element and keep it visible for a pre-determined amount of time. In a confirmation-driven attack, the user would see the border, but because the adversary changes the attack element value back to the original, the content of the user interface element would appear as expected. Understanding that something malicious happened may not be easy for the user. 

\textbf{Change rate limiting.} The user interface could limit rate at which the values of the user interface input elements can be changed. However, our study results show that the majority of the users do not notice even relatively slow UI manipulations that take 250 milliseconds. Finding a rate limit that efficiently prevents user interface manipulation attacks, but does not prevent legitimate user interactions can be challenging. 

\textbf{Randomized user interfaces.} Another way to address our attacks is to randomize parts of the safety-critical system user interface. Both our state tracking algorithm and the attack launch techniques assume a static model of the target system user interface. If the user input elements change their location for every execution, the system state tracking becomes significantly harder. Also attack launch can be complicated by using randomized element locations. Randomized user interfaces have been proposed for smartphone screen lock to prevent shoulder surfing and smudge attacks \cite{lockdown, vonZezschwitz-iui13}. The Intel IPT technology randomizes PIN input to prevent malware from stealing it \cite{intel-ipt}.

While UI randomization can complicate, or even prevent, our attacks, it also increases the chances of human error. In contrast to smartphone screen lock, on safety-critical terminals an increased error rate is typically not acceptable. For example, medical device evaluations consider lack of UI consistency a critical safety violation \cite{graham-2004}. Randomization can increase attack resistance, and thus improve safety, but at the same time incerase human errors, and thus decrease safety. Finding the optimum is an interesting direction for further research, but outside the scope of this paper.

\textbf{Human user tests.} Passwords are often used to authenticate that the correct user is interacting with a computing system. Passwords do not protect against our attacks, because the adversary can learn any entered passwords. CAPTCHAs are a common technique to verify that the user input originates from a human. In our scenario, the attack device cannot solve a CAPTCHA, as it cannot read from the screen, and observing the user to solve one test does not help in future tests. 

The terminal user interface can require that the user must solve a human user test to confirm the safety-critical operation. This approach does not prevent element-driven attacks. Once the adversary has detected interaction with the attack elements, it can wait, modify their values, and after that let the user to complete the test in order to confirm the operation. Also confirmation-driven attacks remain possible with this approach. The user can also be asked to solve a test to be allowed to modify the attack elements. This prevents confirmation-driven attacks. When the user chooses to confirm the safety-critical operation, the adversary cannot return to the attack elements and modify their values without user involvement. For element-driven attacks the adversary has to adjust his attack strategy. The adversary must perform the modification when the user interacts with one of the attack elements (and not shortly after it). Table~\ref{tab:captcha} summarizes these options. 

Human user tests can improve attack resistance, but forcing the user to solve such a test for every modification of a safety-critical UI element is not be acceptable in many systems we consider. 

\begin{table}[t]
  \centering
  \footnotesize
  \begin{tabular}{lcc}
	\hline
	\textbf{Attack type} & \multicolumn{2}{c}{\textbf{CAPTCHA placement}} \\
	& Confirmation & Element mod. \\
	\hline
	confirmation-driven & attack possible & attack prevented \\
	element-driven & attack possible & visibility increased\\
    \hline
  \end{tabular}
  \caption{Placement options for human user tests.}
  \label{tab:captcha}
\end{table}

\textbf{Continuous user authentication.} While traditional user authentication systems require the user to log in once, continuous authentication systems monitor user input over a period of time to detect if the observed usage deviates from a previously recorded user profile. Many such systems track mouse velocity, acceleration and movement direction \cite{Pusara2004, Shen2009, CaiSG14}, together with click events \cite{Shen2009,CaiSG14}, angle-based curvature metrics and click-pause measurements \cite{Zheng2011}. Typically these systems collect user input events for a fixed period of time and then analyze the input to detect unauthorized usage.

The proposed systems that demonstrate low false rejection rates, typically require a significant number of consecutive impostor actions (e.g., 20 consecutive mouse clicks \cite{Zheng2011} or 70 consecutive mouse actions \cite{Mondal:2015}). Even when tailored for higher false rejection rates, the systems need to observe the impostor for significant amount of time (e.g., 12 consecutive seconds \cite{Shen-2013}). Our attacks require only brief mouse movement and one or few clicks, and the attacks can be performed well under a second. Our state estimation works fully passively. Thus, the current continuous authentication systems are not directly applicable to detection of our attacks.

\textbf{Summary.} We conclude that all the reviewed countermeasures have major limitations. Finding better protective measures that are both effective and practical to deploy remains an open problem.

\section{Discussion}
\label{sec:discussion}

\textbf{User study validity.} To evaluate the performance of our state tracking algorithm, we collected user traces online. We also evaluated attack detection online. Our study participants were not domain experts and they had no prior experience on the tested UI. Thus, their observed behavior may not match the one of the real terminal user. Testing our attacks in a real scenario is not safe. However, a controlled lab study on a real terminal and domain experts would provide more confidence to our results. 

\textbf{User interface complexity.} We experimented our attacks on a user interface that consists of approximately ten states. We consider this typical UI complexity for embedded dedicated-purpose terminals. If our approach were to be used on applications with more complex UIs, the state tracking process becomes more challenging and computationally intensive, as the amount of uncertainty and the number of state trackers increase. Evaluating our attack approach on more complex UIs would be an interesting direction for further work. Our focus is on embedded terminals and simple UIs.

\textbf{Non-deterministic user interfaces.} In this work we focus on deterministic interfaces where the same element click always transits to the same, pre-defined UI state. However, embedded terminals can have more complex UIs where this is not the case. For example, clicking a button can lead to different states, depending on the values in the current state. Our attack can be extended to encompass such cases as well. 
 

\textbf{Building the attack device.} In our proof-of-concept implementation, the state estimation and the attack components were decoupled, and were implemented and evaluated separately. We envision a real attack device to be a small, constrained dongle with, e.g., the processing capabilities of a low-end modern smartphone. Even though our tracking algorithm was implemented in unoptimized Python code, the processing delay on our setup was very low (under 50ms). The processing delay on the real attack device would likely be higher, however, as long as the processing delay is shorter than the time needed for a user to perform two consecutive state transitions, our algorithm can track the user in real time. We analyzed all our traces and we found that 95\% of inter-click times are larger than 1000ms, which gives ample time to run an optimized version of the algorithm even on constrained devices. Furthermore, the attack device only needs to track all possible states until it becomes fairly certain about the current state. The device can then discard all but one (the most likely) tracker, and restart tracking --- the processing delay does not accumulate. The attack launching would incur little computational requirements on the attack device.

\section{Related Work}
\label{sec:related}

\textbf{USB attacks.} Key loggers are small devices that the adversary can attach between a keyboard and the target system. The key logger records user input and the adversary collects the device back later to learn any entered user secrets such as passwords. Such attacks are limited to passive information leakage, while our approach enables active runtime attacks with severe safety implications.

A malicious user input device, or a smartphone that impersonates one \cite{wang-acsac10}, can attack PC platforms by executing pre-programmed attack sequences \cite{chen-blackhat09, crenshaw-2011, maskiewicz-woot14}. For example, a malicious keyboard can issue dedicated key sequence to open a terminal and execute malicious system commands. The input device might also be able to copy malicious code to the target system. Such attacks are typically not possible on hardened embedded terminals where the user cannot escape the application UI, and installation and execution of unsigned code is prevented.

\textbf{Terminal protection.} Power analysis can be used to identify unknown (malicious) software processes running on embedded terminals, such as medical devices \cite{clark-healthtech13}. Such approaches would not detect our attack where no malicious code is running on the embedded terminal.

\textbf{User interface attacks.} In systems where multiple applications or websites share the same display, the user can be tricked to interact with false UI elements. For example, a malicious website may be able to draw an overlay over a button that causes the user click the button unintentionally. Such attacks are called clickjacking \cite{huang-usenix12} or UI redressing \cite{niemietz-2011}. In our attack scenario, the adversary can only modify and injects user events. 


	

\section{Conclusions}
\label{sec:conclusion}

In this paper we have presented a new way to attack safety-critical, hardened terminals. In the attack, the adversary installs an attack device between a user input device and the terminal, and the attack is launched when the authorized user is performing a safety-critical operation, by modifying or injecting new user input events. Our approach is easy to deploy on the location, invisible to traditional malware detection, and difficult for the user to notice. Many of the attack variants we tested had success rate over 90\%. We analyzed several countermeasures and noticed that all of them have limitations. We conclude that our attack presents a serious threat to many safety-critical terminals.

{\small
\bibliographystyle{abbrv} 
\bibliography{bibliography}}

\appendix

\section{Supplementary Material}
\label{sec:supplementary}

\textbf{Instructions text.} Your task is to program a pacemaker with the correct settings. The programmer is already connected to the pacemaker. 

\begin{enumerate}
	\item Find the data of the patient which was diagnosed with atherosclerosis.
	\item By clicking the ``Program Pacemaker'' button, program the pacemaker with the settings (``Threshold'', ``Amplitude'', ``Rate'') of that patient.
	\item Click ``Complete Task ''. 
\end{enumerate}

\begin{table}[h]
  \centering
  \small
  \begin{tabular}{lr}
    \hline    
	\textbf{Age} \\
	18-29 & 40.86\% \\
	30-39 & 41.19\% \\
	40-49 & 12.52\% \\
	50-59 & 3.95\% \\
	60 or above & 1.48\% \\
	\hline
  \end{tabular}
  \quad
  \begin{tabular}{lr}
    \hline
    \textbf{Gender} \\
	 Male & 74.14\% \\
	 Female & 25.86\% \\
    \hline
    \textbf{Education} \\
	 Primary school & 2.47\% \\
	 High school & 31.63\% \\
	 Bachelor & 65.90\% \\
    \hline
  \end{tabular}
  \caption{User trace collection demographics.}
  \label{tab:trace-demo}
\end{table}

\begin{table}[h]
  \centering
  \small
  \begin{tabular}{lr}
    \hline    
	\textbf{Age} \\
	18-29 & 41.64\% \\
	30-39 & 38.84\% \\
	40-49 & 13.51\% \\
	50-59 & 4.20\% \\
	60 or above & 1.80\% \\
	\hline
  \end{tabular}
  \quad
  \begin{tabular}{lr}
    \hline
    \textbf{Gender} \\
	 Male & 69.87\% \\
	 Female & 30.13\% \\
    \hline
    \textbf{Education} \\
	 Primary school & 2.10\% \\
	 High school & 30.03\% \\
	 Bachelor & 67.87\% \\
    \hline
  \end{tabular}
  \caption{Attack detection study demographics.}
  \label{tab:attack-demo}
\end{table}

\end{document}